# COCCIA*LAB*

*To discover the causes of social, economic and technological change*

**CocciaLab Working Paper 2019 – No. 39/2019**

**A new technological system to revolutionize the management of valvular heart diseases: transcatheter aortic valve for the treatment of aortic stenosis**

Mario COCCIA

CNR -- National Research Council of Italy
&
Yale University

# A NEW TECHNOLOGICAL SYSTEM TO REVOLUTIONIZE THE MANAGEMENT OF VALVULAR HEART DISEASES: TRANSCATHETER AORTIC VALVE FOR THE TREATMENT OF AORTIC STENOSIS

*Mario Coccia[1]*

CNR -- NATIONAL RESEARCH COUNCIL OF ITALY & YALE UNIVERSITY

CNR -- NATIONAL RESEARCH COUNCIL OF ITALY
Collegio Carlo Alberto, Via Real Collegio, 30-10024 Moncalieri (Torino, Italy)

YALE SCHOOL OF MEDICINE
310 Cedar Street, Lauder Hall, New Haven, CT 06510, USA
*E*-mail: mario.coccia@cnr.it

Mario Coccia ORCID: http://orcid.org/0000-0003-1957-6731

## Abstract

The goal of this study is a technology analysis of a revolution for the management of valvular heart diseases given by new technological system based on transcatheter artificial aortic valve, a delivery catheter and a loading system for the treatment of aortic stenosis (a narrowing of the aortic valve because of calcium build up that restricts blood flow to aorta, the body's main artery) in patients at intermediate and greater risk for surgical therapy. In particular, this study analyzes the technological trajectories of new technology of transcatheter aortic valve implantation (TAVI) in relation to the established technology of surgical aortic valve replacement (SAVR). To explore the emerging technology of TAVI for valvular heart diseases, a model is applied to show how new TAVI technology is substituting established technologies of SAVR to revolutionize the management of aortic stenosis in cardiology. Statistical analyses reveal the sharp increase of the scientific production of TAVI that has a coefficient of growth as function of time equal to 0.40 ($p<.001$) versus a coefficient for SAVR technology of 0.10 ($p<.001$). This finding suggests a technological forecasting that TAVI will be the dominant technology for the treatment of patients with valvular heart diseases. Overall, then, this study shows that new technology of transcatheter aortic valve implantation (TAVI) introduced in 2002 is growing in terms of scientific, inventive and innovative activity to revolutionize the management of aortic stenosis in cardiology worldwide.

## JEL codes: I10; I18; O30; O31; O32; O33; O34

---

[1] Acknowledgement. I gratefully acknowledge financial support from National Research Council of Italy–Direzione Generale Relazioni Internazionali for funding this research project developed at Yale University in 2019 (grant-cnr n. 62489-2018). The author declares that he has no relevant or material financial interests that relate to the research discussed in this paper.

Coccia M. (2019) *A new technological system to revolutionize the management of valvular heart diseases: transcatheter aortic valve for the treatment of aortic stenosis*

**GOALS OF THE INVESTIGATION**

This paper has two goals. The first is to analyze the evolution of new technology for valvular heart diseases based on transcatheter aortic valve implantation (TAVI) in comparison with established technology of surgical aortic valve replacement (SAVR) to treat aortic stenosis[2]. The second goal is to suggest properties that explain and generalize characteristics of technological change in clinical practice of cardiology for improving the management of severe aortic stenosis in society.

This study is part of a large body of research on the evolution of technology to explain characteristics and properties of technological, economic and social change (Coccia, 2017, 2019, 2019a). In the research field of technological evolution, Hosler (1994, p. 3, original italics) argues that the development of technology is, at least to some extent, influenced by "technical *choices*", which express economic, social and political factors, and "technical *requirements*", imposed by material properties. Arthur and Polak (2006, p. 23) claim that: "technology … evolves by constructing new devices and methods from ones that previously exist, and in turn offering these as possible components—building blocks—for the construction of further new devices and elements". In the context of technological evolution, Pistorius and Utterback (1997) argue that technical change can be also due to a rivalry between technologies in competitive markets in which emerging technologies often substitute for more mature technologies (cf., Grodal, 2015; Coccia, 2019b).

This study focuses on the evolution of a new medical technology in cardiology to treat Aortic Stenosis (AS) that is one of the most common valvular heart diseases in society. AS is a narrowing of the aortic valve opening that affects mainly old people and this disease may increase social problems and mortality in society having demographic trends directed to ageing population (Lindroos et al., 1993; Iung et al., 2005). In fact, severe AS of people, without aortic valve replacement, it can lead to an increase of mortality in society. The traditional treatment in the management of severe aortic stenosis for patients is surgical aortic valve replacement (SAVR). However, an emerging technology, alternative to SAVR, is transcatheter aortic valve implantation (TAVI) that is, more and more, a routinely performed technique in cardiology worldwide (Sedeek et al., 2019).

---

[2] *Abbreviations in the text*: TAVI=transcatheter aortic valve implantation; SAVR=surgical aortic valve replacement (SAVR)

Although several contributions in these fields of research, the behavior and characteristics of new medical technology, considering scientific and innovative production, which is generating industrial and corporate in clinical practice are hardly known. In particular, this study addresses some basic questions:

- what are the *degree* and *rate* at which new technology of TAVI is adopted when it attempts in substituting for existing technology of SAVR?
- What are the *evolutionary properties* of TAVI in a setting of competition between technologies in clinical practice of cardiology?
- And finally, what are the *economic consequences* of new technology TAVI for management of AS in health institutions?

Next sections endeavor to explain how emerging technology of TAVI substitutes SAVR, generating a revolution in the management of severe aortic stenosis in society.

## THEORETICAL FRAMEWORK

Arthur (2009, p. 15ff) claims that one of the most important problems to understand in studies of technology is to explain how technological innovation evolves and generates social change (cf., Arthur and Polak, 2006; Basalla, 1988)[3]. Technological evolution can be explained with theories based on processes of competitive substitution of a new technology for the old one (Coccia, 2019b; Fisher and Pry, 1971; Sahal, 1981). In particular, theories of competitive substitution between technologies show that the adoption of a new technology is associated with the nature of some comparable older technology in use (Sahal, 1981; Utterback et al., 2019). This study focuses on competition between medical technologies in the research field of cardiology to treat Aortic Stenosis (AS) that is

---

[3] For other studies of the sources, evolution and transfer of technology see Coccia, 1999, 2005, 2007, 2008, 2012, 2014, 2015, 2015a, 2015b, 2016, 2017a, 2017b, 2018a, 2018b, 2018c, 2019c; Coccia and Wang, 2015.

3 | P a g eCoccia M. (2019) *A new technological system to revolutionize the management of valvular heart diseases: transcatheter aortic valve for the treatment of aortic stenosis*

*CocciaLab Working Paper 2019 – No. 39/2019*

one of the most frequent cardiac problems in society with ageing population. Surgical aortic valve replacement (SAVR) is the traditional technique for patients with severe AS. However, the treatment option of SAVR is associated with a high risk of operative morbidity or mortality in many patients (Armoiry et al., 2018; Lemor et al., 2019). Transcatheter aortic valve implantation (TAVI) is a less invasive medical technology, alternative to conventional technique of surgical therapy (Paparella et al., 2019). Cribier et al. (2002) showed, for the first time in medicine, the feasibility of a percutaneous valve implantation, called TAVI in a patient with AS, providing a promising less invasive alternative treatment to SAVR technique for treating valvular heart diseases. Since 2002, this new medical technology of TAVI is growing with incremental and radical innovations that broaden its applications in clinical practice of cardiology (Bourantas and Serruys, 2014). In fact, technological advances in artificial heart valves with new generations of product innovation have improved the medical technology of TAVI, reduced the risk of complications, allowed the treatment of complex problems in cardiology, increased the efficacy of this technique that is the best treatment option for inoperable patients and high-risk populations as well as recent studies also show potential extension in intermediate-risk patients with symptomatic aortic stenosis (cf., Martinez et al., 2019; Mäkikallio et al. 2019; Thourani et al., 2016). This new medical technology is associated with the evolution of artificial heart valves, delivery catheters and loading systems, such as innovative products by Edwards Lifescience corporation, a worldwide leader in prosthetic valve for surgical market (Edwards Lifescience Corporation, 2019, 2019a, 2019b; Fig. 1).

The history of this technology shows that one of the initial prostheses is the innovative product called Cribier–Edwards, produced in 2003 and consisted of a stainless-steel frame with equine pericardium valve leaflets. In 2006, it is introduced the transcatheter heart valve Edwards SAPIEN using a higher sealing cuff and bovine pericardium leaflets. In 2009, it is introduced SAPIEN XT, which consisted of cobalt chromium alloy frame and bovine pericardium leaflets. The SAPIEN 3 (S3) is the latest generation of Edwards balloon-expandable valves.

Rheude et al. (2018) describe its technical characteristics:

> It features a cobalt chromium alloy frame that provides a high radial strength for circularity and optimal hemodynamics, a low frame height and an open cell geometry, allowing access to coronary vessels for future interventions and an outer polyethylene terephthalate (PET) skirt to minimize paravalvular leakage (PVL). The valve tissue consists of three leaflets manufactured from bovine pericardium. Four different sizes of the S3 THV are currently available: 20mm, 23mm, 26mm and 29mm. Selection of the appropriate THV should be made according to multislice computed tomography (MSCT) annulus area-based sizing recommendations provided by the manufacturer. The treatable range of aortic annulus diameters is wide and ranges from 18.6 mm to 29.5 mm.

| 2003 | 2006 | 2009 | 2018-2019 |
|---|---|---|---|
| *Cribier Edwards SAPIEN* | *Edwards SAPIEN* | *Edwards SAPIEN XT* | *Edwards SAPIEN 3 and SAPIEN 3 Ultra* |
| stainless-steel frame with equine pericardium valve leaflets | higher sealing cuff and bovine pericardium leaflets | cobalt chromium alloy frame and bovine pericardium leaflets | • cobalt chromium alloy frame <br> • an outer polyethylene terephthalate <br> • three leaflets manufactured from bovine pericardium <br> • Four different sizes 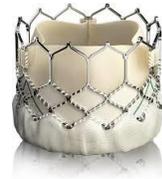 |

**Figure 1**. Evolution of transcatheter heart valves by American medical equipment company Edwards Lifesciences Corporation (2019, 2019a, 2019b)

Another market leader in this industry is the American enterprise Medtronic that in 2009 buys the start-up CoreValve, giving a boost to the design of the prosthesis and the carrier catheter with new innovative products, such as CoreValve, Evolut R and Evolut PRO system (Medtronic, 2019) that is approved by U.S. Food and Drug Administration in 2019.

U.S. Food & Drug (2019) states that:

> The Medtronic CoreValve Evolut R System and Medtronic CoreValve Evolut PRO System each consists of a transcatheter aortic valve (TAV), a delivery catheter, and a loading system. The TAV is an artificial heart valve made of pig tissue attached to a flexible, self-expanding nickel-titanium (Nitinol) frame for support. The Medtronic CoreValve Evolut R System and Medtronic CoreValve Evolut PRO System were previously approved for the treatment of severe aortic stenosis (a narrowing of the aortic valve that restricts blood flow to aorta, the body's main artery) in patients at intermediate and greater risk for surgical therapy. This approval expands the indications for use to patients at low risk for surgical therapy…. The doctor compresses the TAV and mounts it on the end of a tube-like device called a delivery catheter. The TAV is then inserted into the body through an artery in the leg, the artery in the neck, or through a small cut between the ribs. The valve is then released from the catheter; it expands on its own, and anchors to the diseased valve. Once the new valve is in place, it functions the same as the old valve, opening and closing like a door to force the blood to flow in the correct direction… The CoreValve Evolut R and CoreValve Evolut PRO TAVs are used in patients whose own aortic heart valve is diseased due to calcium build up, which causes the valve to narrow (aortic stenosis) and restricts blood flow through the valve. As the heart works harder to pump enough blood through the smaller opening, it eventually becomes weak. This can lead to symptoms and life-threatening heart problems such as fainting, chest pain, heart failure, irregular heart rhythms (arrhythmias), or cardiac arrest. Once symptoms of severe aortic stenosis occur, over half of patients die within two (2) years if the diseased valve is not replaced. …The CoreValve Evolut R and CoreValve Evolut PRO TAVs should only be used in patients who are considered appropriate for transcatheter heart valve replacement therapy by their heart team (including a surgeon)….The CoreValve Evolut R and CoreValve Evolut PRO TAVs can improve blood flow in patients with aortic stenosis. In the clinical study, the two TAVs were shown to be reasonably safe and effective for treating patients with severe aortic stenosis without the need for open-heart surgery. The risk of death or disabling stroke at 2 years was similar in patients receiving a CoreValve Evolut R or CoreValve Evolut PRO TAV and those receiving open-heart surgery….Any procedure to replace the aortic valve carries the risk for serious complications. The serious complications associated with implanting a CoreValve Evolut R or CoreValve Evolut PRO TAV also carries the risk of serious complications such as death, stroke, acute kidney injury, heart attack, bleeding, and the need for a permanent pacemaker. For some patients with coexisting conditions or diseases, the risks may be especially high.

Moreover, the Evolut™ PRO system combines exceptional valve design and advanced sealing with an excellent safety profile (Figure 2). The Evolut PRO system features an external tissue wrap added to the proven platform design. In particular, new design features provide the following sealing mechanisms:

- The external wrap increases surface contact with native anatomy, providing advanced sealing
- The frame oversizing and cell geometry provide consistent radial force across the treatable annulus range

Moreover, the TAVI technology of Evolut PRO system with supra-annular and self-expanding design provides the following benefits:

- Less workload for the ventricle
- Fewer instances of prosthesis-patient mismatch, which has been correlated to improved long-term survival[2]
- A large effective orifice area (EOA) provides improved flow, less resistance, and better long-term durability

Finally, supra-annular valve design maximizes leaflet coaptation and promotes single-digit gradients and large EOAs.

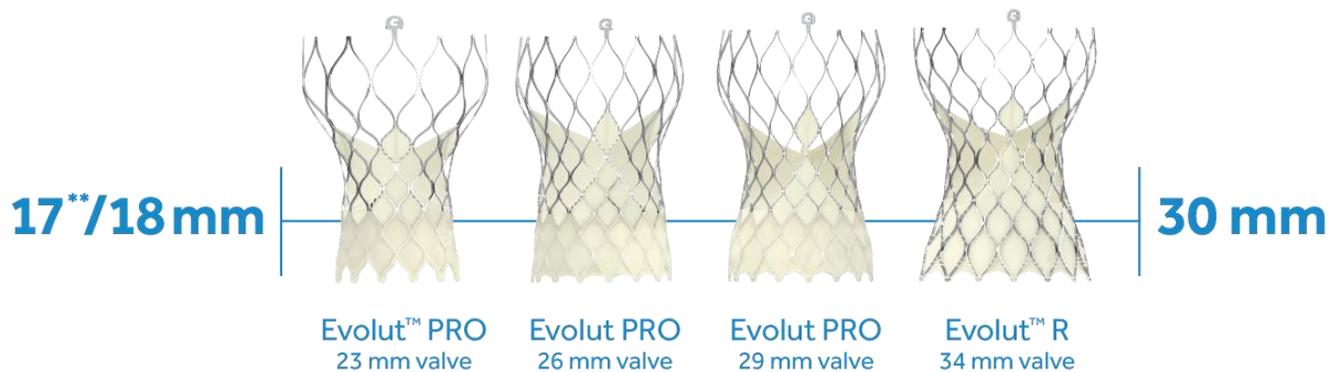

**Figure 2.** Medtronic CoreValve Evolut PRO System for patients whose aortic heart valve is diseased due to calcium build up, which causes the valve to narrow (**aortic stenosis**) and restricts blood flow through the valve. *Source*: Medtronic (2019)

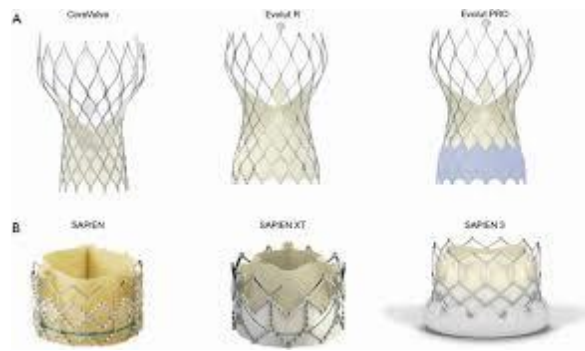

**Figure 3**. Technological evolution of architecture design between artificial aortic valves
Evolut (A) and SAPIEN (B)

Figure 3 shows a comparison of the technological evolution of product architecture in artificial aortic valve SAPIEN by Edwards Lifesciences Corporation and Evolut by Medtronic. Saleem et al. (2019) argue that new technology of TAVI is comparable to SAVR in terms of short-term and mid-term mortality and neurologic events in low surgical risk patients. In this context, Ando et al. (2019) find out that emerging technology of TAVI has a significantly lower composite of all-cause mortality or disabling/major stroke at 1 year compared with treatment of SAVR in low-to-intermediate surgical risk cohort. Lemor et al. (2019) show that in patients greater than 80 years of age, new technology of TAVI is an effective and safer alternative to SAVR since it is associated with lower in-hospital mortality, lower major in-hospital complications, lower 30-day readmission rate and hospital costs (cf., Elbadawi et al., 2019). Hence, since 2002, after the first-in-man case, new technology of transcatheter aortic valve implantation (TAVI) is generating a change of technological paradigm in clinical practice of cardiology with a revolution for the management of severe aortic stenosis, such that is a technique more and more applied in cardiac departments of hospitals worldwide.

## MATERIALS AND METHODS

*Model to analyze technological evolution of medical technology in valvular heart diseases*

The proposed model of technological growth here analyzes a new radical technology (*Kl*) in relation to an established technology *V*. This approach is based on the biological principle of allometry that was originated in zoology to study the differential growth rates of the parts of a living organism's body in relation to the whole body (cf., Reeve and Huxley, 1945). Sahal (1981) suggests this model to explain patterns of the diffusion of technology, providing interesting case studies in the agriculture, manufacturing, steel production, electricity generation, etc.

The model is based on following assumptions.

(1) *V* is established technology, such as SAVR and *Kl* is new technology, such as TAVI.

(2) Let *Kl(t)* be the level of a new technology *Kl* at the time *t* and *V(t)* be the level of an established technology *V* at the same time.

Suppose that both *Kl* and *V* evolve according to some *S*-shaped pattern of technological growth, such a pattern can be represented analytically in terms of the differential equation of logistic function. For *V*, established technology, the starting equation is:

$$\frac{1}{V}\frac{dV}{dt} = \frac{b_1}{K_1}(K_1 - V)$$

The equation can be rewritten as:

$$\frac{K_1}{V}\frac{1}{(K_1 - V)}dV = b_1 dt$$

The integral of this equation is:

$$\log V - \log(K_1 - V) = A + b_1 t$$

$$\log\frac{K_1 - V}{V} = a_1 - b_1 t$$

$$V = \frac{K_1}{1 + \exp(a_1 - b_1 t)}$$

$a_1 = b_1 t$ and $t =$ abscissa of the point of inflection.

The growth of *V(t)* can be described respectively as:

$$\log\frac{K_1 - V}{V} = a_1 - b_1 t \quad [1]$$

*Mutatis mutandis*, for new technology *Kl(t)* the equation is:

$$\log\frac{K_2 - Kl}{Kl} = a_2 - b_2 t \quad [2]$$

The logistic curve here is a symmetrical *S*-shaped curve with a point of inflection at 0.5K, with $a_{1,2}$ that are constants depending on initial conditions, $K_{1,2}$ that are the equilibrium levels of growth and $b_{1,2}$ that are rate-of-growth parameters (1= established technology: i.e., *V*; 2= new technology: i.e., *Kl*).

Solving equations [1] and [2] for $t$, the result is:

$$t = \frac{a_1}{b_1} - \frac{1}{b_1}\log\frac{K_1 - V}{V} = \frac{a_2}{b_2} - \frac{1}{b_2}\log\frac{K_2 - Kl}{Kl}$$

The expression generated is:

$$\frac{V}{K_1 - V} = C_1 \left(\frac{Kl}{K_2 - Kl}\right)^{\frac{b_1}{b_2}} \qquad [3]$$

Equation [3] of evolutionary growth of new technology ($Kl$) in relation to established technology ($V$) in a simplified form, with some mathematical transformations, is given by:

$$Kl = A\,(V)^B \qquad [4]$$

where $A = \dfrac{K_2}{(K_1)^{\frac{b_2}{b_1}}} C_1 \quad$ and $\quad B = \dfrac{b_2}{b_1}$

The logarithmic form of equation [4] is a simple linear relationship:

$$\log Kl = \log A + B\,\log V \qquad [5]$$

$B$ is the coefficient of growth that measures the evolution of new technology $Kl$ in relation to established technology $V$.

This simple model of the evolution of new technology [5] has linear parameters that are estimated with the Ordinary Least-Squares Method. The value of $B$ in the model [5] measures the relative growth of $Kl$ in relation to the growth of $V$ and it indicates different patterns of technological evolution in markets.

In particular,

- $B < 1$, whether new technology $Kl$ has a lower relative rate of change than old technology over the course of time (*under-development of new technology*)

- $B$ has a unit value: $B = 1$, then the new technology $Kl$ substitutes established technology at a proportional rate

of change (*proportional growth of new technology*)

- $B > 1$, whether new technology $Kl$ has greater relative rate of change established technology over the course of time (*development of new technology*)

In short, the coefficient of growth $B$ in the proposed model can be a metric for analyzing the behavior of growth of a new technology, such as TAVI, in relation to an established technology, such as SAVR, in cardiology.

- *Data and their sources*

The empirical analysis is based on data of ScienceDirect (2019) and its tool of Advanced Search to find scientific products that have in title, abstract or keyword the following terms:

- "transcatheter aortic valve implantation (TAVI)"

- "surgical aortic valve replacement (SAVR)"

Scientific products are a main proxy to assess the patters of technology in cardiology to treat aortic stenosis.

- *Measures*

– For SAVR, number of articles and all scientific products in this scientific and technology field (articles, conference papers, conference reviews, book chapters, short surveys, letters, etc.), 1948-2018 period.

– For TAVI, number of articles and all scientific products in this scientific and technology field (articles, conference papers, conference reviews, book chapters, short surveys, letters, etc.), 1993-2018 period.

The evolution of TAVI and SAVR, measured with the number of articles and other scientific products, is important for a comparative analysis of the technological trajectories of different technologies to treat aortic stenosis in society.

- *Model and data analysis procedure*

Model [5] is specified as follows:

$$Log\ Kl_t = log a + B\ log\ V_t + u_t \qquad [6]$$

$a$ is a constant; *log* has base $e = 2.7182818$; $t$=time; $u_t$ = error term

$Kl_t$ is a measure of the scientific growth of new technology TAVI

$V_t$ is a measure of the scientific growth of established technology SAVR in clinical practice

This study also analyzes the evolution of these technologies TAVI and SAVR considering the scientific production in these research fields as a function of time on a semilogarithmic graph.

Model is specified as follows:

$$Log\ spy = \lambda_0 + \lambda_1\ t + \varepsilon_t \qquad [7]$$

*spy* is scientific production in the research field *y* (i.e., TAVI or SAVR)

$\lambda_0$ is a constant

$\lambda_1$ is the coefficient of regression

$\varepsilon_t$ is error term

From this model, we also generate predicted values, residuals, and prediction intervals.

Finally, the technological forecasting of TAVI and SAVR technologies is performed as follows: the procedure selects *Time* as independent variable, whereas dependent variable is scientific production of TAVI and SAVR, and then we use all cases to predict values using the prediction from estimation period through last case in the SPSS Statistics software. Results of technological forecasting are also represented with sequence chart, using the natural logarithm of predicted values from linear model of scientific production of TAVI and SAVR technology as function of time.

The relationships for technology analysis and technological forecasting are investigated using ordinary least squares (OLS) method for estimating the unknown parameters in a linear regression model. Statistical analyses are performed with the Statistics Software SPSS® version 24.

# RESULTS

Abbreviations in table and figures are: TAVI=transcatheter aortic valve implantation; SAVR=surgical aortic valve replacement (SAVR).

**Table 1**   Descriptive statistics

|  | SAVR | LNSAVR | TAVI | LNTAVI |
|---|---|---|---|---|
| N Valid | 60 | 60 | 22 | 22 |
| N Missing | 11 | 11 | 49 | 49 |
| Mean | 258.93 | 4.37 | 507.14 | 3.99 |
| Std. Deviation | 349.32 | 1.91 | 687.66 | 2.80 |
| Skewness | 1.76 | −0.62 | 0.99 | 0.00 |
| Kurtosis | 2.23 | −0.25 | −0.69 | −1.68 |

*Note*: TAVI=transcatheter aortic valve implantation; SAVR=surgical aortic valve replacement (SAVR)

**Table 2**   Parametric estimates of the model of technological substitution of TAVI on SAVR

*Dependent variable*: *log* scientific products concerning TAVI (new technology)

|  | Constant $\log \alpha$ (St. Err.) | Coefficient B (St. Err.) | $R^2$ adj. (St. Err. of the Estimate) | F (sign.) |
|---|---|---|---|---|
| TAVI | −22.84*** | 4.32*** | 0.95 | 373.92 |
|  | (1.39) | (0.22) | (0.65) | (0.001) |

*Note*: *** significant at 1‰; Explanatory variable is *log* scientific products concerning SAVR (established technology)

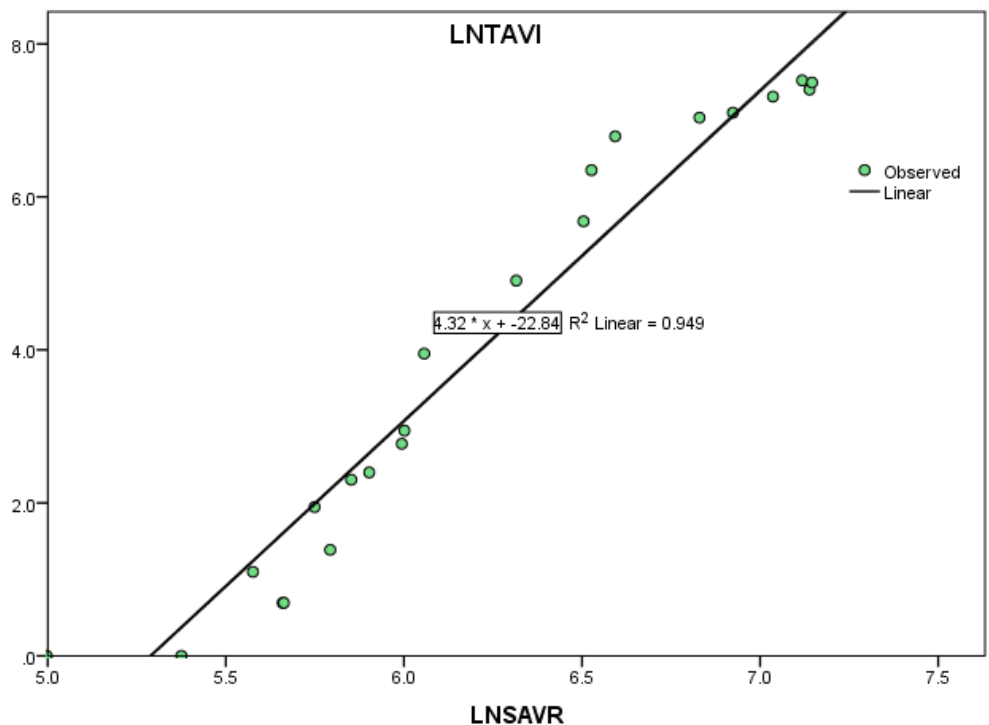

**Figure 4.** Fit line of the growth of TAVI on SAVR (log scale)

Table 1 shows descriptive statistics and the normality of distribution of variables under study based on skewness and kurtosis coefficients. The estimated relationship of proposed model [6] in Tab. 2 and represented in Fig. 4 with the line of best fit shows that the significance of coefficients and explanatory power of equation are high. In fact, the coefficient $R^2$ adj. is high and model of TAVI (new technology) on SAVR (established technology) explains more than 94% variance in the data (Tab. 2). The results show that new technology of TAVI has $B= 4.3$ (i.e., B>1) such that it is substituting with a high relative rate of change the established technical procedure of SAVR over the course of time.

**Table 3** Parametric estimates of the model of technology growth of SAVR as a function of time

| | Constant $\lambda_0$ (St. Err.) | Coefficient $\lambda_1$ (St. Err.) | $R^2$ adj. (St. Err. of the Estimate) | F (sign.) |
|---|---|---|---|---|
| SAVR | −202.58*** | 0.104*** | 0.95 | 1019.0 |
| | (6.48) | (0.003) | (0.45) | (0.001) |

*Note*: *** significant at 1‰; Explanatory variable is *time in years*.

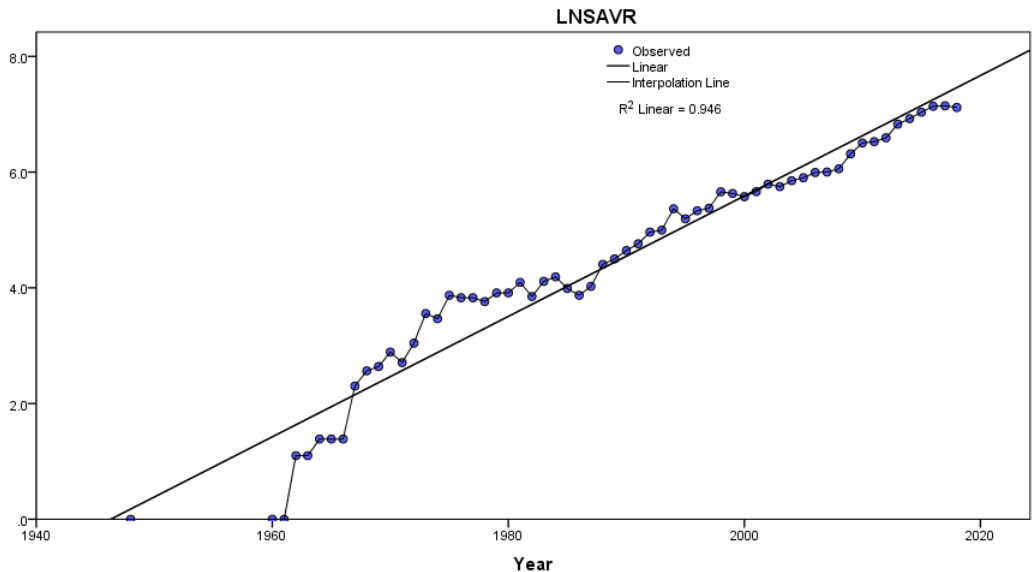

**Figure 5.** Fit line of the growth of SAVR on years

**Table 4** Parametric estimates of the model of technology growth of TAVI as a function of time

*Dependent variable*: *log* scientific products concerning TAVI (new technology)

| | Constant $\lambda_0$ (St. Err.) | Coefficient $\lambda_1$ (St. Err.) | $R^2$ adj. (St. Err. of the Estimate) | F (sign.) |
|---|---|---|---|---|
| TAVI | −777.16*** | 0.39*** | 0.94 | 322.07 |
| | (43.53) | (0.022) | (0.69) | (0.001) |

*Note*: *** significant at 1‰; Explanatory variable is *time (in years)*

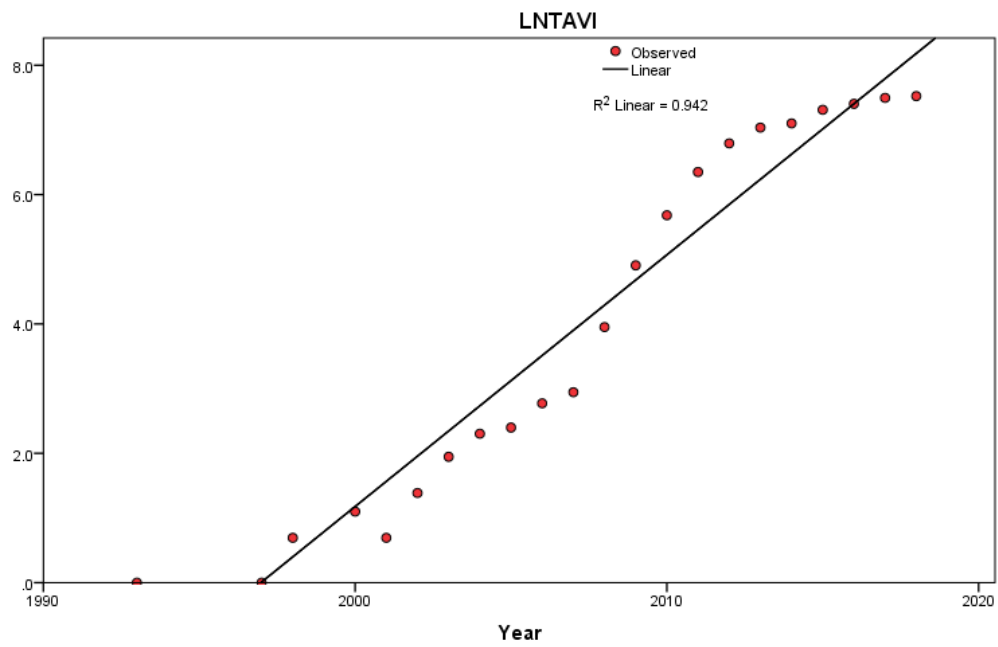

**Figure 6.** Fit line of the growth of TAVI on years

Tab. 3-4 and Fig. 5-6 show results of estimated relationships and trends of SAVR and TAVI as function of time and a comparison of trend over the course of time in figure 7. Results reveal that the scientific and technological production of SAVR is increasing with a coefficient of regression equal to 0.10 ($p<0.001$, semi-log scale), whereas new technology of TAVI is growing over time with a coefficient of 0.40 ($p<0.001$). These findings suggest that transcatheter aortic valve implantation (TAVI) has a higher acceleration than SAVR, such that it can revolutionize the management of severe aortic stenosis with the potential of becoming a dominant medical technology and technique in clinical practice to treat AS in cardiac centers worldwide.

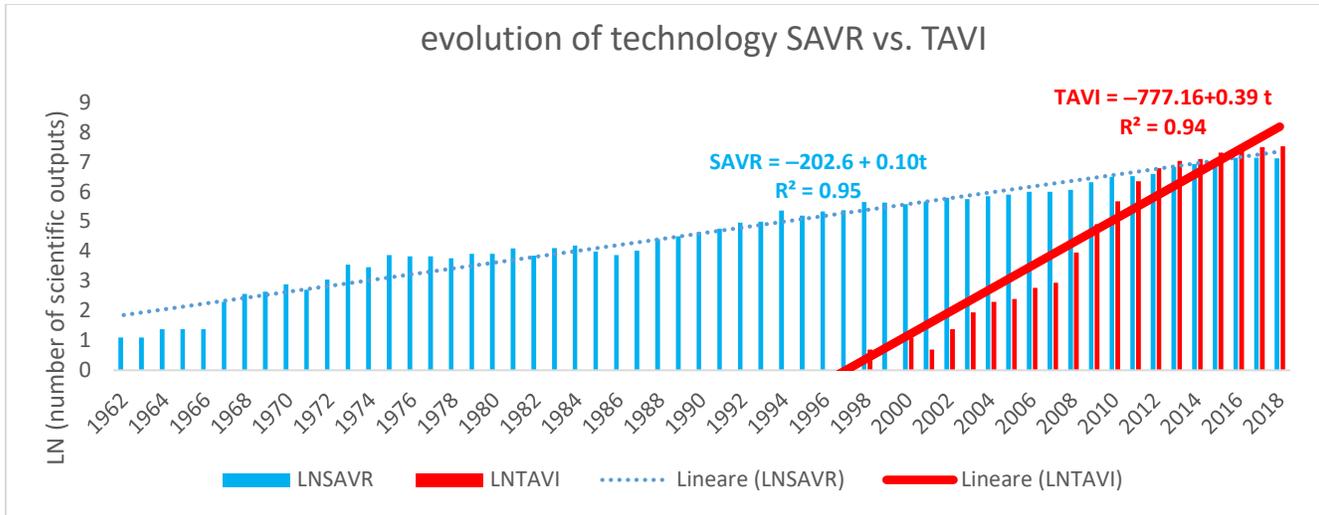

**Figure 7.** Comparison of trends and lines of best fit of the growth between TAVI and SAVR as function of time (years)

Technological forecasting produces the following results for SAVR (table 5 and figure 8).

**Table 5.** Parametric estimates of technology forecasting of SAVR technology

| | *Constant* $\alpha$ *(St. Err.)* | *Case sequence Coefficient* $\beta$ *(St. Err.)* | $R^2$ *adj. (St. Err. of the Estimate)* | *F (sign.)* |
|---|---|---|---|---|
| SAVR | −405.52*** | 16.08*** | 0.67 | 119.95 |
| | (65.99) | (1.47) | (201.14) | (0.001) |

*Note*: *** significant at 1‰; Explanatory variable is *time*. Predict from estimation period through last case, it predicts values for all cases, based on the cases in the estimation period in SPSS software.

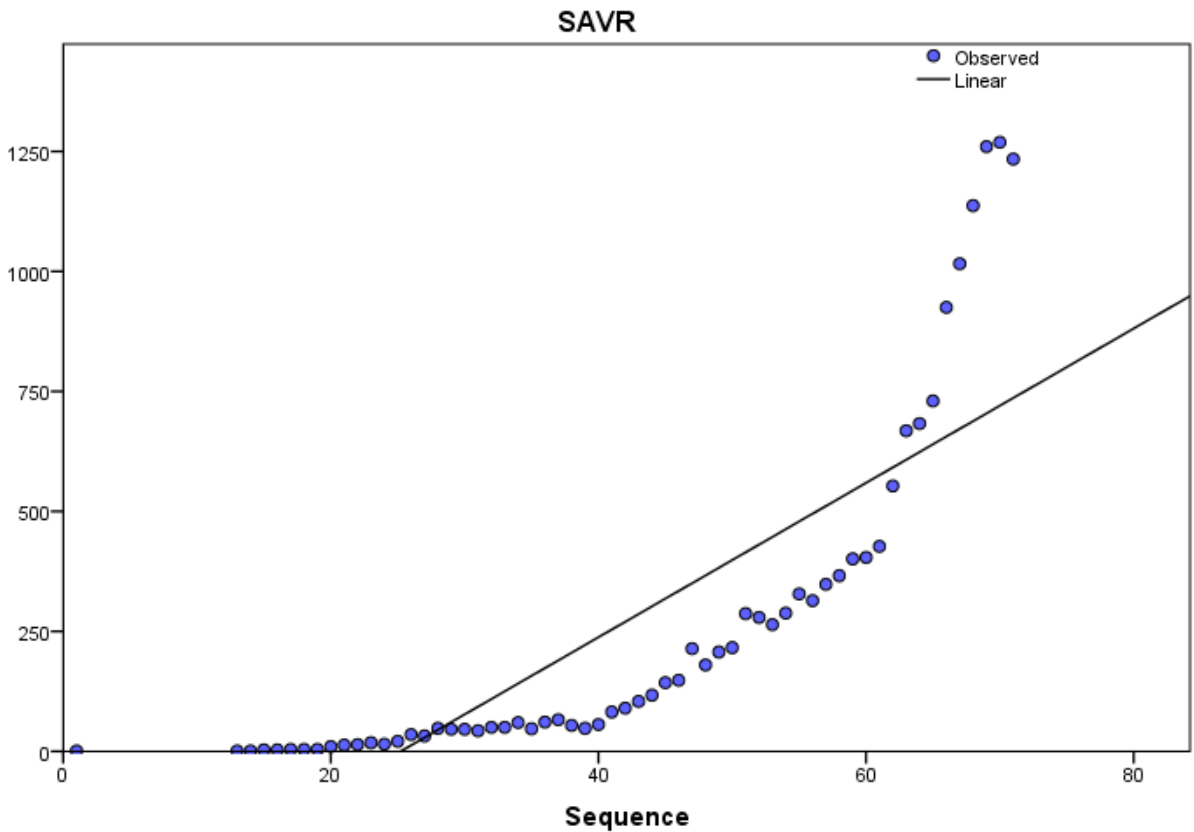

**Figure 8.** Technological forecasting of SAVR technology.
*Note.* Predict from estimation period through last case, it predicts values for all cases, based on the cases in the estimation period in SPSS software.

Technological forecasting for new technology of TAVI is in table 6 and figure 9.

**Table 6.** Parametric estimates of technology forecasting of TAVI technology

| *Dependent variable*: scientific production of TAVI (new technology), using predict from estimation period through last case | | | | |
|---|---|---|---|---|
| | *Constant* $\alpha$ *(St. Err.)* | *Case sequence Coefficient* $\beta$ *(St. Err.)* | $R^2$ *adj.* *(St. Err. of the Estimate)* | *F* *(sign.)* |
| TAVI | −4539.39*** | 83.79*** | 0.71 | 52.07 |
| | (703.83) | (11.61) | (371.20) | (0.001) |

*Note*: *** significant at 1‰; Explanatory variable is *time*; Predict from estimation period through last case, it predicts values for all cases, based on the cases in the estimation period in SPSS software.

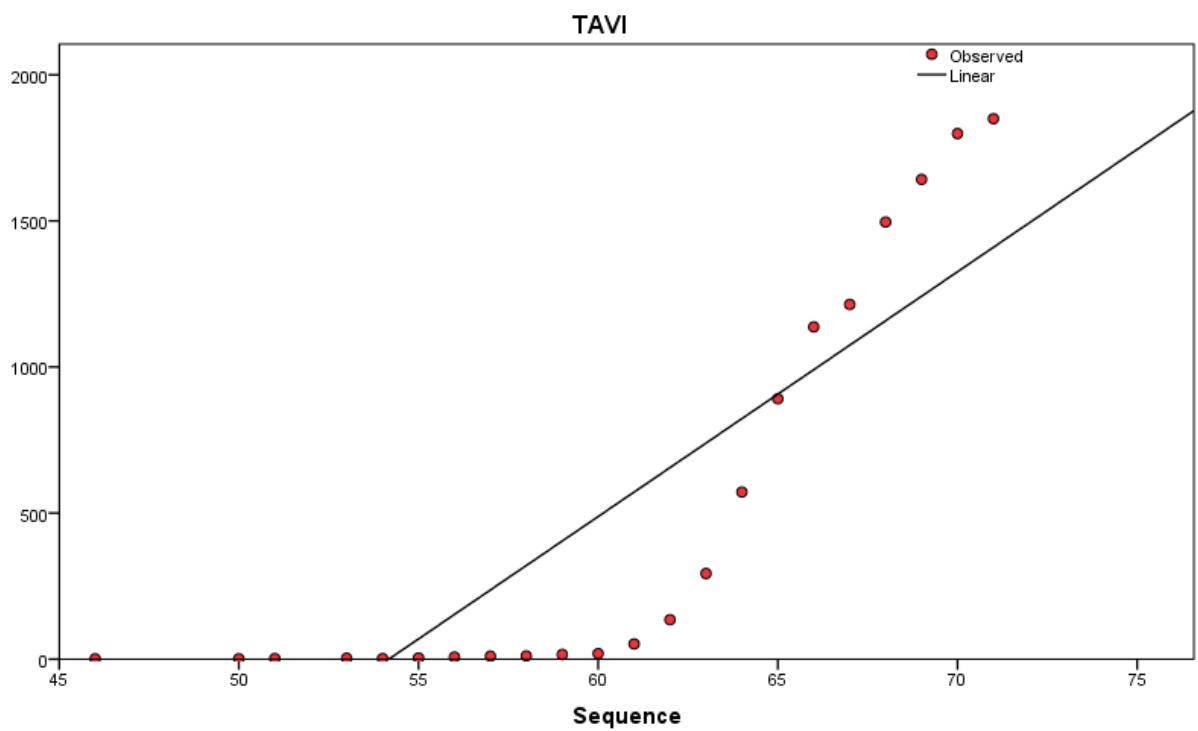

**Figure 9.** Technological forecasting of TAVI technology
*Note.* Predict from estimation period through last case, it predicts values for all cases, based on the cases in the estimation period in SPSS software.

The comparative analysis of the coefficient of regression based on case sequence shows that SAVR technology has a forecasted magnitude of growth equal to 16.08, whereas TAVI has a predicted level of growth of 83.79. These results further suggest a technological forecasting in cardiology oriented towards a scientific and technological growth of TAVI technology in clinical practice.

Using the sequence chart of predicted values and prediction interval at 95% of the scientific production for TAVI and SAVR, in natural logarithm, based on curve fit of linear model, results reveal that SAVR is a technology in maturity phase, whereas new technology of TAVI is in the technological phase of growth (Figure 10).

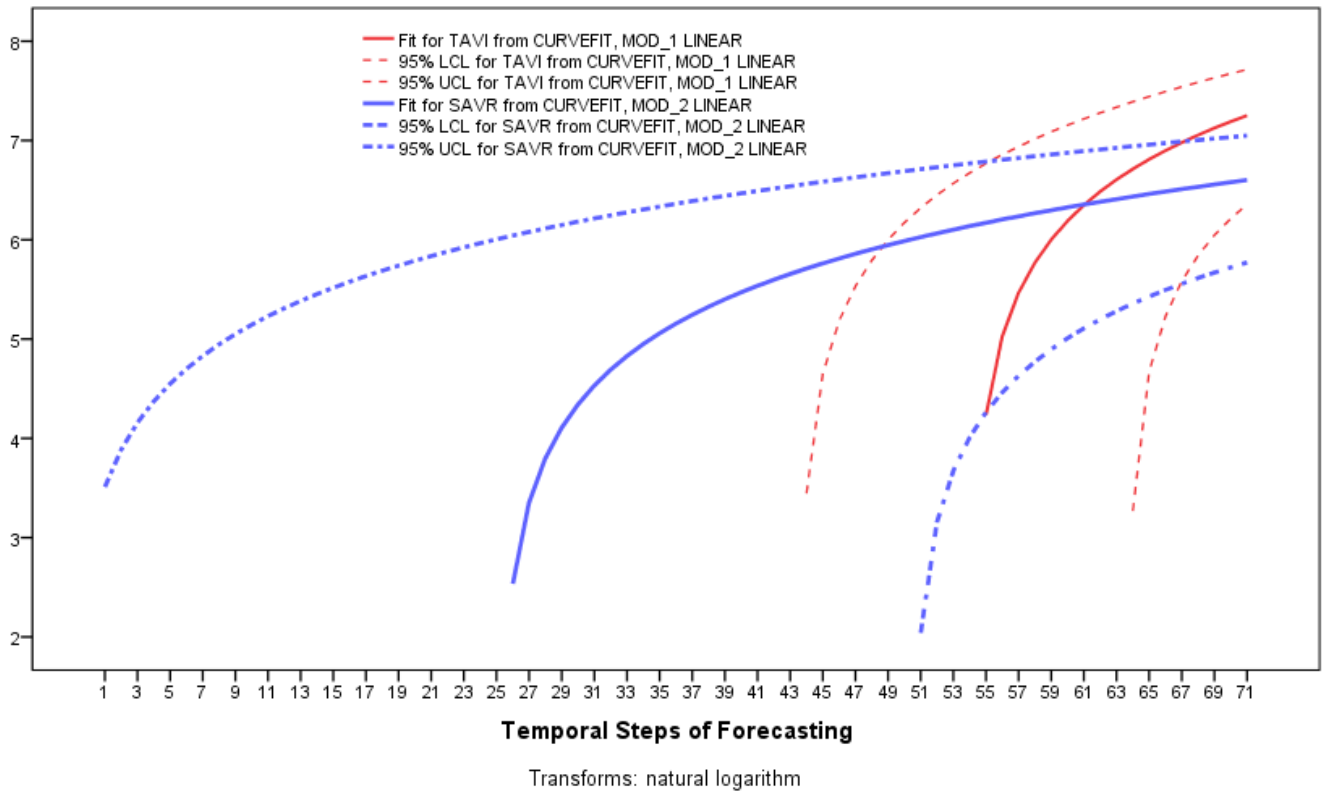

**Figure 10.** Technological forecasting of TAVI vs. SAVR using sequence plot with predicted values and prediction intervals (95% upper and lower bounds from curve fit of linear model of regression) on y-axis (semilog scale)

The findings suggest that TAVI technologies have scientific and technological growth higher than established SAVR technologies. This scientific and technological growth of new technology TAVI in cardiology can be due to ambidexterity learning processes, given by:

− "*learning via diffusion*" (Sahal, 1981, p. 114, Italics added) in which the increased adoption of TAVI technology supports the path for improvement in its technical characteristics (i.e., technological advances).

− "*diffusion by learning*" that improvement in the technical characteristics of TAVI technology enhances the scope for its adoption over the course of time (cf., Sahal, 1981, p. 114).

**DISCUSSION AND LIMITATIONS**

The concept of competition is frequently used to explain the diffusion and evolution of innovation and technology in industrial economics (Fisher and Pry, 1971; Porter, 1980; Utterback et al., 2019). The competition between technologies leads to a process of disruptive creation that generates industrial and corporate change over time (Calvano, 2006; Coccia, 2018; Coccia, 2019b). In particular, when comparable technologies do exist, each technology tends to affect the behavior and evolutionary pathway of other technologies (Coccia, 2019, 2019b). Pistorius and Utterback (1997) argue that emerging technologies often supplant for more mature technologies in markets. The dynamics between technologies is usually referred to as competition that leads to the dominance of a new technology on another one in turbulent markets (cf., Berg et al., 2019). Fisher and Pry (1971, p. 88) state that: "The speed with which a substitution takes place is not a simple measure of the pace of technical advance . . . . it is, rather a measure of the unbalance in these factors between the competitive elements of the substitution". In general, competition is often embodied in substitutes, which have a powerful force in markets to improve products and processes and generate technical, economic and social change. Porter (1980) considers substitutes as one of the five forces of industrial competition.

In cardiology, TAVI technology is generating a process of actual substitution for the old SAVR technique and, as a consequence, technical, industrial and corporate change in medical sectors.

Results of empirical analysis here suggest some properties and predictions:

1. The nature, significance and evolution of TAVI technology is always associated with some comparable established technology, such as SAVR

2. The growth of TAVI technology is generally an allometric process of growth given by a disproportionate growth of new TAVI technology in relation to the old SAVR technology in cardiology

3. In the short run, new technology of TAVI can induce incremental technological advances of established technology SAVR that has a prospect of being supplanted by new technology TAVI in cardiology
4. In the long run, new technology of TAVI has a series of technological advances of its own resulting from various major and minor innovations to pave the way to be the dominant technology over other established technologies in cardiology to treat AS.
5. The ambidexterity learning processes, based on *learning via diffusion* and *diffusion by learning,* are driving forces underlying the development and adoption of new technology TAVI in complex and fast changing field of research, considering low-, intermediate-, high-surgical risk of patients.

This study may be basic in modern economic systems of rising healthcare costs, because policy-makers and clinicians must make, more and more, difficult decisions regarding resource allocation to increase organizational efficiency and improve health of patients, reducing related risks. New technology TAVI has reimbursement that, in 2010s, is restricted to high-risk and inoperable patients only. However, Ando et al. (2019) argue that medical technology TAVI has significantly lower all-cause mortality or disabling/major stroke and disabling/major stroke compared with SAVR also in low-to-intermediate surgical risk patients at 1 year. Zhou et al. (2019) argue that TAVI in intermediate-risk patients has a higher immediate procedural costs than SAVR, driven primarily by the cost of the transcatheter valve (about $33,000) but this was offset by a shorter length of hospitalization following TAVI, such that the combined cost of initial procedure and hospitalization in TAVI was lower than SAVR. In general, medical technology TAVI is likely to be highly cost-effective compared to SAVR in intermediate-risk patients with severe aortic stenosis (AS). In particular, TAVI using the third-generation SAPIEN 3 and Evolut PRO system would be cost-effective compared to SAVR in treatment of intermediate-risk patients with severe symptomatic AS. Moreover, over a ten-year horizon, medical technology SAPIEN 3 TAVI is associated with greater quality adjusted life expectancy and lower long-term costs compared with SAVR technology. Goodall et al. (2019) also show that in intermediate risk patients, new technology of TAVI is associated with superior clinical outcomes compared to

surgery and is cost saving too. Hence, it could be expected that cost savings are conservative and likely to increase over time. In fact, Mäkikallio et al. (2019) claim that technological evolution of TAVI has led to reduced rates of perioperative bleeding, stroke, severe acute kidney injury and other procedural complications placing medical technology TAVI as an effective standardized technique with safe and feasible treatment results for most patients with symptomatic severe AS. The longer survival of TAVI operations, the reduction of major and minor complications as well as the decreased demand for postoperative care will likely cause a higher TAVI implantation rate in the future among patients with severe AS and not only. In fact, the prediction is that about 200,000 annual candidates will be treated with medical technology of TAVI in 2020 (Durko et al., 2018; O'Sullivan and Wenaweser, 2017).

This new technology of TAVI needs further incremental innovations however. In fact, several concerns remain for medical technology of TAVI. One of the technical issues is the valve durability. Blackman et al. (2019) report a long-term structural valve degeneration post-TAVI (median 5.8 years) from the United Kingdom registry that showed 8.7% of moderate and 0.4% of severe valve degeneration. Instead, Søndergaard et al. (2019) argue that structural valve deterioration is significantly higher in SAVR at 6 years. Another concern is the subclinical leaflet thrombosis that has been reported to be higher in TAVI than SAVR and possibly associated with worse outcomes. Ando et al. (2019) also show that stroke remains to be a major perioperative issue both TAVI and SAVR. Mäkikallio et al. (2019, p. 277) also reveal that several issues will play a key role in expanding TAVI indication in low- and intermediate-risk patients, such as valve durability, need for permanent pacemaker, valve performance as well as optimal antithrombotic or anticoagulation medications are questions needed to be answered when expanding TAVI technology to low-risk patient populations (Terré et al., 2017; cf., Chakos et al., 2017). Another practical issue to be solved is whether TAVI is to replace SAVR among younger patient population because of a relatively high permanent pacemaker implantation rate for some TAVI technological devices (van Rosendael et al., 2018). In short, since the short-term TAVI results in terms of mortality, stroke, valve performance and vascular complications have been excellent, now the main attention is directed to long-term durability of TAVI technological valves and patients with low-intermediate surgical risks. To sum up, Table 6 shows main *pros* and *cons* of medical technology of TAVI in clinical practice.

**Table 6.** *Pros* and *cons* of new medical technology of transcatheter aortic valve implantation

| *Strengths TAVI* | *Weaknesses TAVI for low-risk patients* |
|---|---|
| ▫ Medical technology appropriate for inoperable patients | ▫ Need optimal anticoagulation medications |
| ▫ TAVI is a less invasive alternative to conventional surgical therapy SAVR | ▫ Long-term valve durability |
| ▫ Effective and safer technique | ▫ Valve performance |
| ▫ Lower rate of acute kidney injury | ▫ Need for permanent peacemaker |
| ▫ Lower rate of bleeding complications, blood transfusions, cardiac tamponade | ▫ Readmission rates for sepsis and gastrointestinal bleeding |
| ▫ Reduced rates of perioperative bleeding and stroke | ▫ TAVI among younger patients because of a relatively high permanent pacemaker implantation rate for some TAVI devices |
| ▫ Fewer hospital day and likely to be discharged home with cost saving | |
| ▫ Lower rates of in-hospital complications | |
| ▫ Cost saving over time | |
| ▫ TAVI has significantly lower all-cause mortality or disabling/major stroke and disabling/major stroke compared with SAVR in low-to-intermediate surgical risk patients at 1 year. | |

In general, the study suggests how the competition between technologies generates technological and industrial change in markets, such as in cardiology. These results here show that competition between SAVR and TAVI technology is generating a corporate change in health institutions for the management of aortic stenosis. Findings here can support innovation strategy of hospitals on critical decisions of when to invest in R&D of new TAVI technologies, abandon the old SAVR technology or pursue an intermediate level of R&D investment between old and new technology for sustaining and safeguarding cost-effectiveness of medical institutions and health of patients.

The study here is a reasonable starting point for understanding behavior and characteristics of these new medical technologies in cardiology. However, we know, *de facto*, that other things are often not equal over time and space in the domain of technology. Overall, then, the study here may lay the foundation for development of more sophisticated analyses at the intersection between economics of innovation and medical economics to explain technological forecasting of new technology and support strategic management of hospitals and research labs. In fact, these findings here can encourage further exploration in the *terra incognita* of the competition between new and established technologies in medicine that generates disruptive creation for technological and social change in society. Future efforts in this research field will be also directed to provide further empirical evidence, also considering dependency-network framework between technologies to better explain the nature and behavior of new technologies in complex organizations, such as medical institutions (cf., Iacopini et al., 2018). To conclude, the study of new technology in medicine with *pros* and *cons* of short-, medium-, long term for institutions and individuals is a non-trivial exercise. In fact, Wright (1997, p. 1562) properly claims that: "In the world of technological change, bounded rationality is the rule."

Coccia M. 2019a. A Theory of classification and evolution of technologies within a Generalized Darwinism, Technology Analysis & Strategic Management, vol. 31, n. 5, pp. 517-531, http://dx.doi.org/10.1080/09537325.2018.1523385

Coccia M. 2019b. Killer Technologies: the destructive creation in the technical change. ArXiv.org e-Print archive, Cornell University, USA. Permanent arXiv available at http://arxiv.org/abs/1907.12406

Coccia M. 2019c. Why do nations produce science advances and new technology? Technology in society, https://doi.org/10.1016/j.techsoc.2019.03.007

Coccia M., Wang L. 2015. Path-breaking directions of nanotechnology-based chemotherapy and molecular cancer therapy, Technological Forecasting & Social Change, vol. 94, pp. 155–169, https://doi.org/10.1016/j.techfore.2014.09.007

Cribier A, Eltchaninoff H, Bash A, Borenstein N, Tron C, Bauer F, Derumeaux G, Anselme F, Laborde F, Leon MB. 2002. Percutaneous transcatheter implantation of an aortic valve prosthesis for calcific aortic stenosis: first human case description. Circulation. 106:3006–3008.

Durko A.P., Osnabrugge R.L., Van Mieghem N.M., et al. 2018. Annual number of candidates for transcatheter aortic valve implantation per country: current estimates and future projections. Eur Heart J.39:2635–2642.

Edwards Lifesciences Corporation 2019. TAVI with the SAPIEN 3 valve https://www.edwards.com/gb/devices/Heart-Valves/Transcatheter-Sapien-3 (Accessed October 2019)

Edwards Lifesciences Corporation. 2019a. Reproduced from https://www.edwards.com [homepage on the Internet]. transcatheter heart valve. Available from: https://www.edwards.com/gb/devices/heart-valves/transcatheter.

Edwards Lifesciences Corporation. 2019b. https://www.edwards.com [homepage on the Internet]. transcatheter heart valve. Available from: https://www.edwards.com/gb/devices/heart-valves/transcatheter.

Elbadawi A., Ahmed H.M.A., Mahmoud K., Mohamed A.H., Barssoum K., Perez C., Mahmoud A., Ogunbayo G.O., Omer M.A., Jneid H., Chatterjee A. 2019. Transcatheter Aortic Valve Implantation Versus Surgical Aortic Valve Replacement in Patients with Rheumatoid Arthritis (from the Nationwide Inpatient Database). American Journal of Cardiology, vol. 124, n. 7, pp. 1099-1105

Fisher J. C., Pry R. H. 1971. A Simple Substitution Model of Technological Change, Technological Forecasting & Social Change, vol. 3, n. 2-3, pp. 75-88.

Goodall G., Lamotte M., Ramos M., Maunoury F., Pejchalova B., de Pouvourville G. 2019. Cost-effectiveness analysis of the SAPIEN 3 TAVI valve compared with surgery in intermediate-risk patients, Journal of Medical Economics, 22:4,289-296, DOI: 10.1080/13696998.2018.1559600To link to this article: https://doi.org/10.1080/13696998.2018.15596

Grodal S., Gotsopoulos A., Suarez F. F. 2015. The coevolution of technologies and categories during industry emergence. Academy of Management Review, vol. 40, n. 3, pp. 423–445. http://dx.doi.org/10.5465/amr.2013.0359

Hosler D. 1994. The Sounds and Colors of Power: The Sacred Metallurgical Technology of Ancient West Mexico. MIT Press, Cambridge.

Iacopini I., Milojević S., Latora V. 2018. Network Dynamics of Innovation Processes, Phys. Rev. Lett., vol. 120, n. 048301, pp. 1-6, https://doi.org/10.1103/PhysRevLett.120.048301

Coccia M. (2019) *A new technological system to revolutionize the management of valvular heart diseases: transcatheter aortic valve for the treatment of aortic stenosis*

*CocciaLab Working Paper 2019 – No. 39/2019*